\documentclass[aps,pra,amsmath,amssymb,twocolumn
]{revtex4-1}

\usepackage[utf8]{inputenc}
\usepackage[english]{babel}
\usepackage[T1]{fontenc}

\usepackage{amsmath,amssymb}
\usepackage{bm}
\usepackage{amsthm}
\usepackage{mathtools}
\usepackage{hyperref}

\usepackage{listings}

\usepackage{graphicx}
\graphicspath{{./},{./figs/}}
\usepackage{xcolor}

\begin{document}
\title{Three-level rate equations in cold, disordered Rydberg gases}
\author{R.~V.~\surname{Skannrup}}
\email{r.v.skannrup@tue.nl}
\author{T.~v~\surname{Weerden}}
\author{Y.~vd~\surname{Werf}}
\altaffiliation[Currently at]{ LaserLaB, Department of Physics and Astronomy, Vrije Universiteit, Amsterdam, The Netherlands}
\author{T.~\surname{Johri}}
\author{E.~J.~D.~\surname{Vredenbregt}}
\author{S.~J.~J.~M.~F.~\surname{Kokkelmans}}
\affiliation{Eindhoven University of Technology. P.O. Box 513, 5600 MB Eindhoven, The Netherlands}
\date{\today}

\begin{abstract}
We have investigated formation of structures of Rydberg atoms excited 
from a disordered gas of ultracold atoms, using rate equations 
for two-photon Rydberg excitation in a single atom without eliminating the 
intermediate state. We have explored the validity range of these rate 
equations and defined a simple measure to determine, 
whether our model is applicable for a given set of laser parameters. 
We have applied these rate equations in Monte Carlo simulations of 
ultracold gases, for different laser beam profiles, and compared these 
simulations to experimental observations and find a general agreement. 
\end{abstract}

\maketitle

\section{Introduction}
\label{sec:intro}
Highly excited atoms, generally referred to as Rydberg atoms, show 
extreme features such as long life times and strong dipole interactions, 
first observed in 1981 \cite{Raimond1981}. 
As a result of these strong interactions, a Rydberg atom will block its 
neighbors from being excited, as the Rydberg level is moved out of 
resonance with the excitation laser. This blockade effect, 
first observed in 2009 \cite{Urban2009}, 
has been proposed \cite{Jaksch2000,Lukin2001} 
as the mechanism for a two qubit quantum gate, specifically a CNOT gate 
first demonstrated in 2010 \cite{Isenhower2010}.
Rydberg atoms have also been proposed as a many-body spin model 
quantum simulator \cite{Weimer2010}, and realized \cite{Labuhn2016}.
In addition, the opposite mechanism, known as facilitation, is also 
possible \cite{Lesanovsky2014,Lesanovsky2013,Valado2016}, 
and is characterized by resonant excitation of Rydberg atoms at specific 
distances from existing Rydberg atoms. 
An in depth review of quantum information with Rydberg atoms is 
available \cite{Saffman2010}. 

Properties of Rydberg ensembles are often studied
through measuring counting statistics such as the Mandel Q-parameter 
and spectra \cite{Mandel1979,Schempp2014,Malossi2014}. 
From these results different phase 
transitions can be recognized, for instance between a facilitation and 
blockade regime, which was already predicted for systems in equilibrium 
\cite{Weimer2008}. 
Another method to study Rydberg atoms is by measuring spatial 
correlations through spatial imaging \cite{Schausz2012,PhysRevA.87.023401}. 
Often the three-level system is 
simplified to a two-level system, which is only possible if a large 
laser detuning is used \cite{Shavitt1980,Brion2007}. 
However, no matter how simple an atom description is, the state space 
grows exponentially in the number of atoms (just like with qubits) and 
one must still find a way to make many-body calculations feasible.
We translate the problem to a Markov process with a limited amount of 
possible transitions, characterized by transition rates, and then employ 
Monte Carlo techniques as done in \cite{Ates2007}.

This research was done with a specific experiment, described in 
\cite{Engelen14,Bijnen14}, in mind, though 
it is not limited to describing this. In our lab in 
Eindhoven University of Technology the setup can trap rubidium atoms in 
a magneto-optical trap (MOT) and excite these to Rydberg states. 
The excitation region can be varied at will by means of a spatial light
modulator (SLM) with good control of both shape and dimensionality. 
The excitation region does not have to be continuous or convex, but we 
will limit the work presented here to one and two dimensional boxes, as 
this is of more general value. 

In order to describe the versatile experimental 
excitation conditions, we develop a Monte Carlo model based on 
three-level atoms capable of covering the range of laser parameters 
and excitation volume geometries available to the experiment. 
The (de-)excitation 
probabilities of the Monte Carlo simulation are based on rate equations, 
where the detunings and Rabi frequencies of both the Rydberg and 
intermediate states are tunable in the model. In addition to the laser 
parameters, also the choice of intermediate and Rydberg state is kept 
free, by having the spontaneous decay rates of both states and Van der 
Waals coefficient of the Rydberg as input parameters. 
The resulting 
single-atom rate equations go beyond a simple effective two-level 
treatment and are applicable to both resonant and 
off-resonant excitations. We have checked the 
validity of the model and set limits to the validity range. 
This general rate equation description of the single atom, dependent 
on the internal states of the surrounding atoms, can then be used to 
describe the entire cloud in our Monte Carlo simulation, which we use 
to explain our experimental observations.

\begin{figure}
\includegraphics
{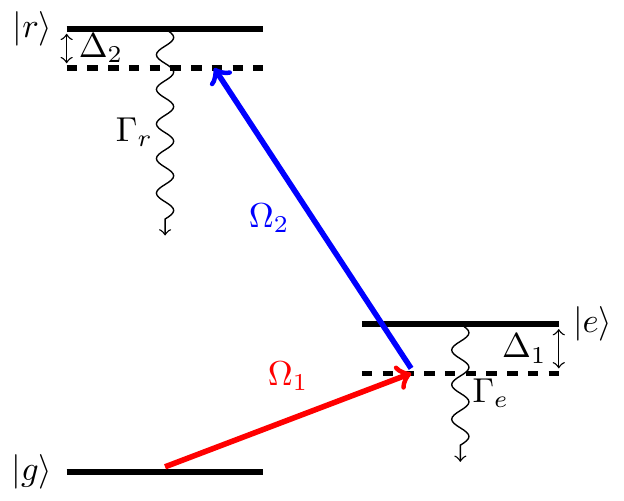}
\caption{\itshape Three level excitation scheme. A red laser (red) excites 
from the ground state $|g\rangle$ to an intermediate state $|e\rangle$ 
and is detuned by $\Delta_1$. Another laser will excite from this 
intermediate level to the Rydberg level $|r\rangle$ and is detuned by 
$\Delta_2$.
\label{fig:two_level}}
\end{figure}

This paper is structured into seven sections. In section \ref{sec:rateeq} 
we will derive the (de-) excitation rates for a single atom influenced 
by lasers as indicated Figure~\ref{fig:two_level}, and in section 
\ref{sec:validity} we investigate the limits to the rate equation model 
(RE) in depth. 
In section \ref{sec:monte} we investigate the differences in a Monte 
Carlo simulation, stemming from the three sets of rates.
In section \ref{sec:conc} we draw conclusions on this work.

\section{Rate equations}
\label{sec:rateeq}
\begin{figure*}
\includegraphics[width=\textwidth]{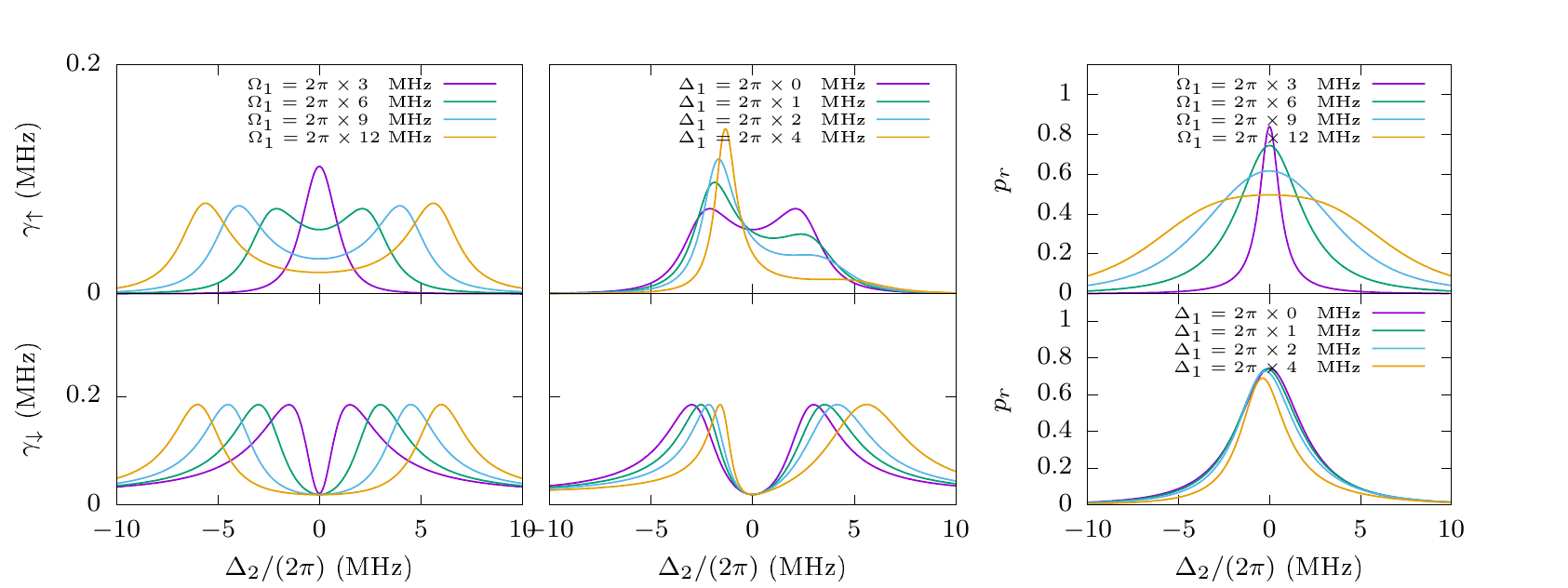}%
\caption{\itshape
Excitation rates (top) and deexcitation rates (bottom) vs blue detuning 
$\Delta_2$, for different sets of laser parameters. 
Left column: Varying red Rabi frequency $\Omega_1$, with constant blue 
Rabi frequency $\Omega_2=2\pi\times 0.4$ MHz and red detuning 
$\Delta_1=2\pi\times 0$ MHz.
Middle column: Varying red detuning $\Delta_1$, with constant blue 
Rabi frequency $\Omega_2=2\pi\times 0.4$ MHz and red Rabi frequency  
$\Omega_1=2\pi\times 6$ MHz.
Right column: Steady state Rydberg populations, while varying 
red Rabi frequency (top) or red detuning (bottom). Other parameters 
are as in the corresponding (de-)excitation column.
\label{fig:rateVar}}
\end{figure*}

We base our approach on $N$ three level atoms with the Hamiltonian 
of the $i^\text{th}$ atom 
in the interaction picture given by 
\begin{align}
H^i=&\frac{\hbar}{2}\bigg(
-\Delta_1|e_i\rangle\langle e_i|
-\Delta_2|r_i\rangle\langle r_i|\nonumber\\&
+\Omega_1|g_i\rangle\langle e_i|
+\Omega_2|e_i\rangle\langle r_i|+ \text{H.C.}\bigg)
\label{eq:ham}
\end{align}
with $|e_i\rangle$ ($|r_i\rangle$) the $i^\text{th}$ atom being in the 
intermediate (Rydberg) state. 
Other authors have used similar rate equation models for two-level 
\cite{Lesanovsky2014,Garttner2014b} or three-level 
\cite{Ates2007PRL,Ates2007,Heeg2012,Ates2011} atoms. 
In Refs.~\cite{Ates2007PRL,Ates2007,Heeg2012} an effective two-level system 
is achieved by fixing the intermediate level detuning at zero. 
Our three-level atom rate equation model allows for both the Rydberg 
and intermediate level detunings to be freely chosen input 
parameters, as well as both Rabi frequencies and the spontaneous 
decay rates. Further, we treat in some detail the limits of this 
free choice in the next section.

In the following sections, the parameters of the three-level atom model will be loosely based on the parameters of our $^{85}$Rb experiment. Then in Section \ref{sec:experiment} we will discuss the validity of this model in the context of a particular experimental realisation. The intermediate state 
is given by $5P_{3/2}$, and for now the Rydberg state is specified as $100S_{1/2}$. 
These states have spontaneous decay rates 
$\Gamma_e=2\pi\times 6.07$ MHz for the intermediate state and
$\Gamma_r=2\pi\times 0.003$ MHz for the Rydberg state.
We call the laser associated with subscript 1 
the probe laser and the one associated with subscript 2 the coupling 
laser. The interactions between atoms are given by
\begin{align}
H^i_\text{int}&=\frac{1}{2}\sum_{\substack{
j\neq i}}^N
\frac{C_6}{r_{ij}^6}|r_j\rangle|r_i\rangle\langle r_i| \langle r_j|,
\end{align}
where $r_{ij}$ is the distance between atoms $i$ and $j$, and $C_6$ 
is the van der Waals coefficient scaling as the principal quantum 
number of the Rydberg state $|r\rangle$ to the power of $11$. 
Since this work is done with a specific ultra cold gas experiment 
in mind, we apply 
the frozen gas approximation and assume the $r_{ij}$s to 
stay constant and the terms of $H_\text{int}^i$ can be evaluated only 
once. The state of the system determines what terms to be included at 
any given time. We will assume that each atom can be modeled independently, 
with only an effective shift in local coupling detuning due to the 
interactions between Rydberg atoms. The entire effect of 
$H_\text{int}^i$  is then captured by modifying the coupling detuning since we consider istropic interactions only:

\begin{align}
\Delta_2 \rightarrow \Delta_2
+\frac{1}{\hbar}\langle \bm{s}| H_\text{int}^i|\bm{s}\rangle 
=\Delta^i_\text{eff}(\bm{s}),
\label{eq:effdel}
\end{align}
where $|\bm{s}\rangle$ is the state of the full $N$ atom system, if 
atom $i$ is excited.

Using these we find the master equation (ME) of atom $i$ in 
Lindblad form
\begin{align}
\frac{d}{dt}\rho=&-\frac{i}{\hbar}[H^i(\bm{s}),\rho]+\mathcal{L}(\rho),
\label{eq:ME}
\end{align}
where $H^i(\bm{s})$ given by eq.~\eqref{eq:ham} with the replacement 
eq.~\eqref{eq:effdel} and Liouvillian 
\begin{align}
\mathcal{L}(\rho)=&
\Gamma_e|e_i\rangle\langle g_i|\rho |g_i\rangle\langle e_i| 
-\frac{\Gamma_e}{2}\{\rho,|e_i\rangle\langle e_i|\} \nonumber\\
+&\Gamma_r|r_i\rangle\langle e_i|\rho |e_i\rangle\langle r_i| 
-\frac{\Gamma_r}{2}\{\rho,|r_i\rangle\langle r_i|\},
\end{align}
and $H^i(\bm{s})$ being the Hamiltonian in eq. \eqref{eq:ham} with 
$\Delta_2$ replaced by the state dependent effective detuning eq. 
\eqref{eq:effdel}.

Rewriting the density matrix of a single atom in vector form, the 
effective time evolution operator is 
\begin{align}
\dot{\bm{\rho}}=\mathcal{R}\bm{\rho}
=\begin{pmatrix}
\mathcal{R}_p & L\\L^T & \mathcal{R}_c
\end{pmatrix}\bm{\rho},
\label{eq:rhodotvec}
\end{align} 
with $\bm{\rho}^T=(\rho_{gg},\rho_{ee},\rho_{rr},\rho_{ge},\rho_{er},
\rho_{gr},\rho_{eg},\rho_{re},\rho_{rg})$, $\mathcal{R}_p$ is a 
$3\times 3$ matrix taking population to populations, $\mathcal{R}_c$ 
is a $6\times 6$ matrix taking coherences to coherences and $L$ is 
a $3\times 6$ matrix taking coherences to populations.
Adiabatically removing the coherences, which is permitted when the 
time scale of the dynamics in the coherences is much smaller than 
the time scale of the population dynamics,
we can write the optical 
Bloch equations
\begin{align}
\dot{\bm{p}}
=\left(\mathcal{R}_p -L\mathcal{R}_c^{-1}L^T\right)\bm{p}=Q\,\bm{p},
\label{eq:vecdev} 
\end{align}
where $\bm{p}=(p_g,p_e,p_r)^T$ are the populations 
($\rho_{gg}$, $\rho_{ee}$ and $\rho_{rr}$) of the 
$i^\text{th}$ atom. Our analysis has shown that, for laser parameters 
where adiabatic elimination of the coherences is valid, the elements of 
$Q$ solely associated 
with the dynamics between $p_g$ and $p_e$ are larger than those 
associated with dynamics of $p_r$ by two or three orders of magnitude.
We will call the terms associated with the dynamics of $p_r$ the 
'small' terms of $Q$.

The general solution to such a homogeneous system of coupled 
differential equations is known
\begin{align}
\bm{p}=\sum_{k} \bm{v}_k \exp\left(-\lambda_k t\right),
\end{align}
with $\lambda_k$ ($\bm{v}_k$) the eigenvalues (-vectors) of 
$Q$, and the sum 
running over all eigenvalues. At sufficiently long time, all but one 
(non-zero) eigenvalues have dampened out, and we know that 
$\dot p_r=\alpha \dot p_e$, where $|\alpha|\gtrsim 1$, since
only one eigenvector contributes to the derivative.
From this we can express $p_e$ in terms of $p_r$ and $Q$
\begin{align}
p_e=&\frac{(0,1,0)\, Q \, (1,0,0)^T}%
{(0,1,0)\,Q\,(-1,1,0)^T}\,(1-p_r),
\label{eq:prape}
\end{align}
by neglecting the small entries in the matrix $Q$.

We define excitation rate $\gamma_\uparrow$ and deexcitation rate 
$\gamma_\downarrow$, such that
\begin{align}
\dot{p}_{r}
=\gamma_\uparrow(p_{g}+p_{e})-\gamma_\downarrow p_{r}
=\gamma_\uparrow(1-p_{r})-\gamma_\downarrow p_{r},
\label{eq:gamdev}
\end{align}
which can be found by substituting eq.~\eqref{eq:prape} into 
eq.~\eqref{eq:vecdev} to get 
\begin{align}
\gamma_\uparrow=&
(0,-\xi,1)\,Q\,
(1,0,0)^T
\label{eq:gamma_up}\\
\gamma_\downarrow=&
(0,0,1)\,Q\,
(\xi-1,-\xi,1)^T.
\label{eq:gamma_down}
\end{align}
with 
\begin{align}
\xi=&\frac{(0,1,0)\,Q\,(-1,0,1)^T}%
{(0,1,0)\,Q\,(-1,1,0)^T}.
\end{align}
For the steady state solution the time derivative is the null-vector, 
and we get
\begin{align}
p_{r}^{\infty}=\frac{\gamma_\uparrow}{\gamma_\uparrow
+\gamma_\downarrow}.
\label{eq:steadystate}
\end{align}

In Fig.~\ref{fig:rateVar} we show the derived (de-) excitation rates 
versus the blue laser detuning $\Delta_2$ for a variety of laser 
parameters. In the left and middle column only a single of 
the three remaining controllable parameters is varied and the others 
are kept constant at $\Omega_1=2\pi\times 6$ MHz, $\Delta_1=0$ MHz and 
$\Omega_2=2\pi\times 0.4$ MHz. In the right most column the 
corresponding steady state populations are shown. 
We observe that the excitation 
rates show the general features we expect, like Autler-Townes splitting 
into two Lorentzian peaks for $\Omega_1>\Gamma_e$ and tend to 0 for 
large Rydberg detuning $\Delta_2$. For small $\Omega_1$, the two 
Lorentzian peaks merge, as expected. For large intermediate state 
detuning $\Delta_1$, the center of the largest peak shifts towards 
larger Rydberg detuning by a value of $\Delta_1/2$ and the 
peak to peak distance is $\sqrt{\Omega_1^2+\Omega_2^2+\Delta_1^2}$. 
Further, the ratio between the maximal height of the two peaks tends 
to zero. These features are easily explained from the eigenvalues 
and -states of the Hamiltonian \eqref{eq:ham}. 
For the de-excitation rates we generally observe the same with two 
important addenda. First, for large Rydberg detuning $\Delta_2$, the 
de-excitation rate does not go to zero, but rather $\Gamma_r$, as 
expected. Secondly, for small Rydberg detuning, the de-exitation 
also approaches the spontaneous decay rate $\Gamma_r$. This happens as 
the de-excitation rate only has an effect on the Rydberg state, which 
for zero detuning has an overlap with a dark state of the 
Hamiltonian \eqref{eq:ham} larger than $0.99$, meaning that the lasers 
have no influence on the de-excitation rate in this case, reducing 
the de-excitation rate to the spontaneous decay rate.

The Rydberg populations, as shown in the rightmost column of 
Fig.~\ref{fig:rateVar}, are found using Eq.~\eqref{eq:steadystate}. 
Increasing the red Rabi frequency $\Omega_1$ beyond the intermediate 
state spontaneous decay rate $\Gamma_e$ broadens and lowers the 
Rydberg transition resonance, as Autler-Townes splitting affect 
the excitation rate. Increasing the intermediate state detuning, 
lowers and narrows the Rydberg transition resonance, but only 
slightly, as well as giving a slight shift to the resonance.

\section{Rate equation validity}
\label{sec:validity}
\begin{figure}
\includegraphics{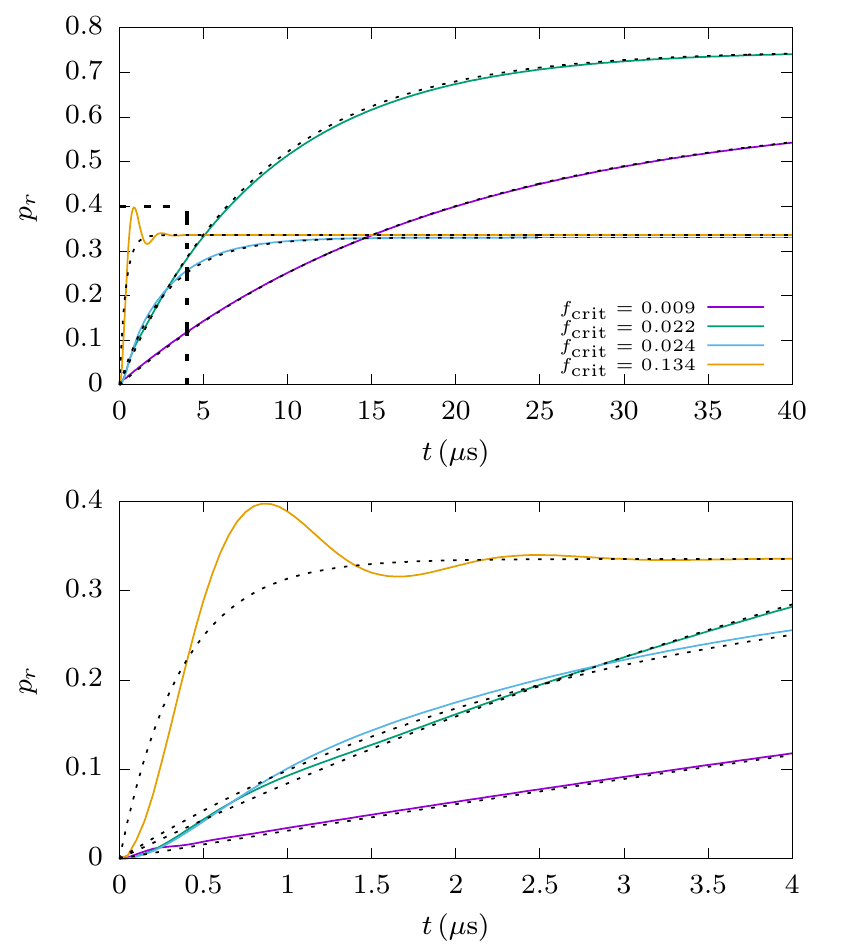}
\caption{\itshape
Numerical solutions to the optical Bloch equation (solid) and rate 
equation approximations (dotted) versus time for different critical 
fractions $f_\text{rm}$.
Bottom plot shows a magnification indicated by dashed square. 
Lines with small critical fraction show good agreement 
between rate equations and optical Bloch equations, whereas the large 
critical fraction line only shows good agreement with final population 
value.
\label{fig:OBEComp}}
\end{figure}
Going back to eq.~\eqref{eq:rhodotvec} and considering the adiabatic 
elimination of the coherences, we know that $\mathcal{R}$, has 
nine complex eigenvalues $v_1,\dots,v_9$ and the coherence only part 
$\mathcal{R}_c$ six complex eigenvalues $k_1,\dots,k_6$, which 
determine the time scale of the dampening of the coherences. 
This time scale has to be short, in order to support adiabatic 
elimination of the coherences.
Of the eigenvalues $v_1,\dots,v_9$, only three have eigenvectors 
with non-zero populations 
of the intermediate state, and of these only one is always real valued, 
we call this $v_1$. In addition one eigenvalue, which we call $v_9$, 
is always zero.

We define the dampening time scales for eigenvalue $v$ as
$\tau_v=\frac{-1}{\Re(v)}$, and make shorthands for the three most 
important 
\begin{align}
\tau_0=&\tau_{v_1}\\
\tau_1=&\max(\tau_{v_2},\dots,\tau_{v_8})\\
\tau_2=&\max(\tau_{k_1},\dots,\tau_{k_6}).
\end{align}
Since $\tau_0$ is, in general, larger than $\tau_1$, it is the time scale 
on which higher order dynamics of the system dampens out and only the 
steady state solution remains. On top of this the coherences dampen out 
on the time scale $\tau_2$. 

\begin{figure}
\includegraphics[width=\columnwidth]{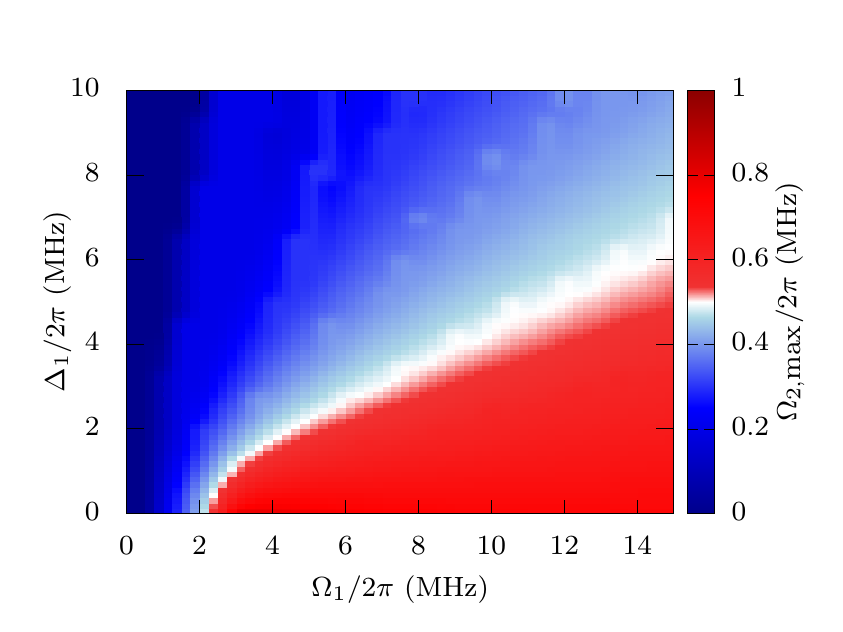}
\includegraphics[width=\columnwidth]{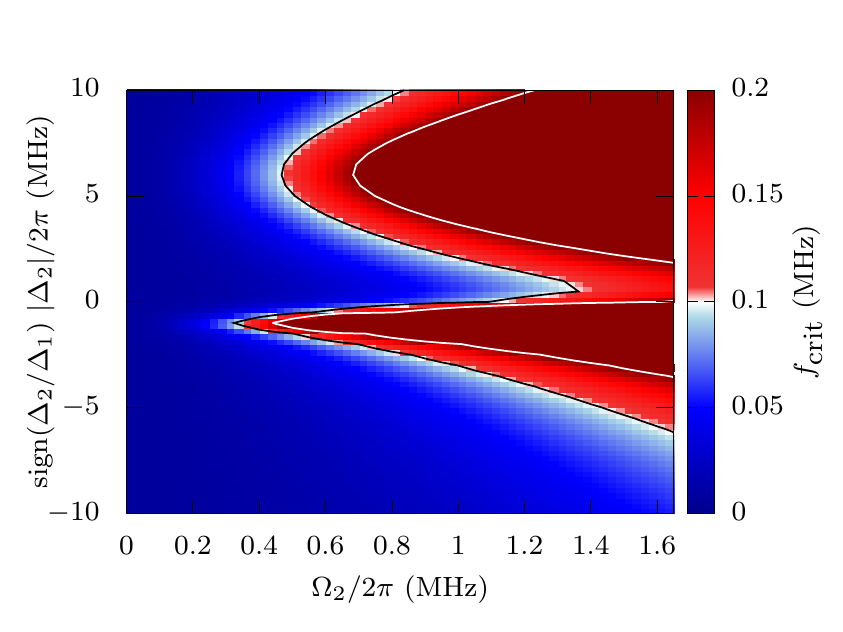}
\caption{\itshape
Top: 
Maximal allowed blue Rabi frequency $\Omega_2$ versus red Rabi 
frequency 
$\Omega_1$ and red detuning $\Delta_1$. This value is found as the 
largest blue Rabi frequency for which the critical fraction is smaller 
than $0.1$ for all (positive) blue detunings $\Delta_2$.
Bottom: 
Critical fraction $f_\textrm{crit}$ in the $\Omega_2$-$\Delta_2$ plane 
for $\Omega_1=2\pi\times 5$ MHz and $\Delta_1=2\pi\times 5$ MHz. 
Black contour line is at $f_\textrm{crit}=0.1$ and white at 
$f_\textrm{crit}=0.2$.We generally require that 
$f_\textrm{crit}<0.1$. Note that the positive detunings allow for 
larger $\Omega_2$, but since the sign is dependent on the ratio between 
the detunings, we can always choose positive for a given experiment.
\label{fig:validity}}
\end{figure}

The rate equation model is dependent on there being a clear distinction 
between the dynamics on the long time scale $\tau_0$ and the shorter 
time scales $\tau_1$ and $\tau_2$, and 
this means that we now have conditions for the 
validity of the rate equation model
\begin{align}
\tau_0>>&\max(\tau_1,\tau_2)\\
t>>&\tau_1,
\end{align}
where $t$ is the simulation run time. 
For practical purposes, we will assume this to be satisfied if
\begin{align}
f_\textrm{crit}=\frac{\max(\tau_1,\tau_2)}{\tau_0}<0.1,
\end{align}
which we call the critical fraction. Fig.~\ref{fig:OBEComp} shows 
a comparison between our rate equation model (dotted) and the optical 
Bloch equations (solid) for different critical fraction values. The 
agreement between solid and dotted lines is clearly dependent on the 
critical fraction. Of special note is the rather bad agreement between 
the solid and dotted lines for critical fraction larger than $0.1$, 
which we choose as the practical limit.

We have searched the parameter space to satisfy these conditions, and 
the requirements are in general quite lax for reasonable red 
laser parameters, see Fig.~\ref{fig:validity}.
For any given combination of red Rabi frequency and detuning, we find 
the critical fraction in the $\Omega_2$-$\Delta_2$ plane, and determine 
the maximal allowed blue Rabi frequency $\Omega_{2,\textrm{max}}$ as
the largest value of $\Omega_2$ for which the critical fraction is smaller 
than $0.1$ for all (positive) blue detunings $\Delta_2$. 

The maximal blue Rabi frequency is dependent on the relative sign of 
blue detuning to red detuning, and we can find blue Rabi frequency 
limits $\Omega_b^+$ and $\Omega_b^-$, dependent on that sign, below 
which the rate equations always hold.
This asymmetry is resulting from the asymmetry 
in the excitation due to the red laser. Since 
$\Omega_b^+\geq\Omega_b^-$ is always the case, we can choose the sign of 
the red detuning such that we have the largest allowed range for 
$\Omega_b$. However, if scanning the blue detuning across the resonance, 
the blue Rabi frequency has to be below $\Omega_b^-$, as the limiting 
point is right below zero. Note that for most 
detunings, the maximal blue Rabi frequency can be much larger than this 
limiting value, and it would be prudent for any experiment to determine 
the limiting values appropriate for the specific experiment.

\section{Monte Carlo simulation}
\label{sec:monte}

We have implemented our (de-)excitation rates in a kinetic Monte Carlo 
simulation, where we extract the values of interest as the average over 
many realizations. We will explore three different settings, 
first we consider a 1D regular lattice, secondly a random gas 
with a square quasi 2D excitation volume and finally we will compare 
to experimental measurements.  The calculations are based on a long-range van der Waals interaction for two $^{85}$Rb atoms in the n=100$S_{1/2}$ state, with a van der Waals coefficient $C_6=56$ THz $\mu$m$^6$ \cite{ARC17}.

\begin{figure}
\centering
\includegraphics[width=\columnwidth]{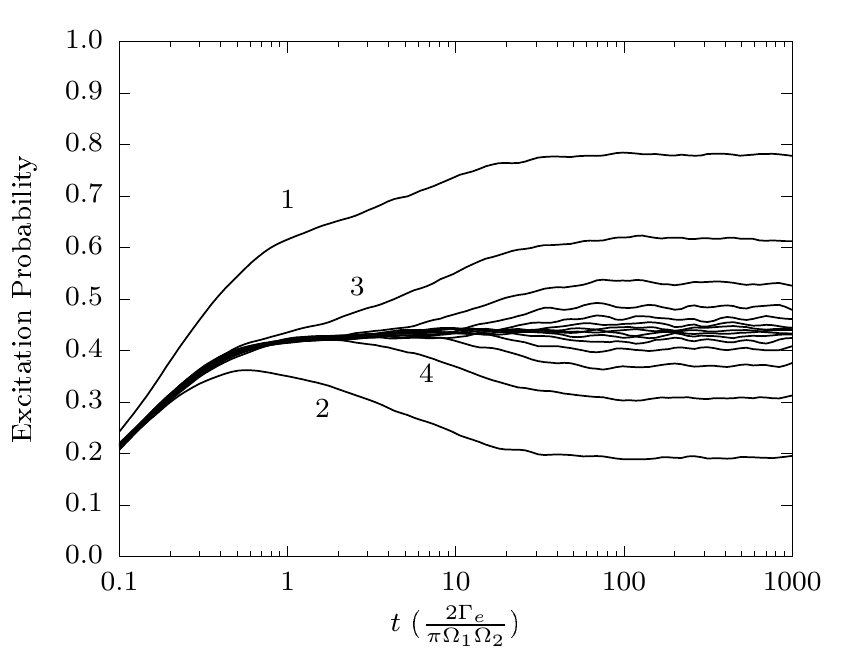}
\caption{\itshape
Time dependence of excitation probability in the 1D lattice with 30 lattice 
sites. Labels denote 
lattice site. Probability is calculated as the number of Monte Carlo 
realizations with a Rydberg excitation at a site divided by the total 
number of realizations for every time step.
\label{fig:mPos}
}
\centering
\includegraphics[width=\columnwidth]{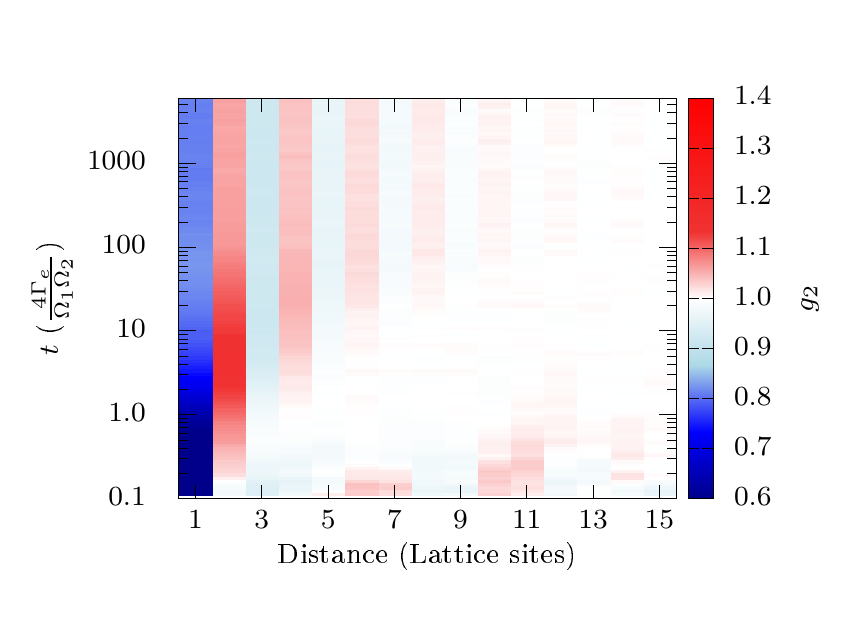}
\caption{\itshape
Time dependence of the $g_2$ function for a 1D lattice with 30 lattice 
sites. Nearest neighbors are never Rydberg pairs, but for times larger
the equilibration time $T=4\Gamma_e/\Omega_1\Omega_2$ 
Rydberg excitations 
start to come in pairs separated by one atom and at $10T$ the crystal 
consists of 3 excitations. The correlation length scales as $t^{1/3}$, 
but saturates at $t\sim 100T$.
\label{fig:g2map}
}
\end{figure}

Each Monte Carlo realization is performed by, at time $t_0=0$, calculating 
(de-)excitation rates for all atoms and an exponentially distributed 
random time step $dt$, with mean value $1/\sum_i \gamma^i_\updownarrow$, 
with $\gamma^i_\updownarrow$ the excitation rate for atom $i$ 
(deexcitation rate if atom $i$ is already excited). 
We then randomly pick an atom with probability proportional to the 
$\gamma^i_\updownarrow$. This atom is then (de-)excited and the time is 
set to $t_1=t_0+dt$. This procedure is then repeated until the time 
exceeds the simulation time $t_\textrm{end}<t_{\textrm{end}-1}+dt$.
At prespecified times, we save the state of the system for our analysis. 
The output of the Monte Carlo simulation is the average over all 
realizations.

For our 1D lattice simulation, the individual lattice sites are 
identifiable and we therefore explore the time evolution of the 
excitation probability of the individual sites. This requires many 
realizations to converge and we therefore perform 6000 realizations for 
the simulation. For a realised population of 0.5, the statistical error bar for this number of realisations is about $2\%$.

We consider a string of 30 atoms placed at regular distance 
$l=16.3\,\mu\textrm{m}$ with laser parameters $\Omega_1=\Gamma_e/2$, 
$\Omega_2=\Gamma_e/10$, $\Delta_1=\Delta_2=0$. This results in a 
nearest neighbor interaction strength $V_{nn}=2\Omega_1$ for the 
$100S$ state and corresponds to the work done in \cite{PhysRevA.87.023401}.
Our work is consistent with their result, but for a slightly larger 
correlation length due to the fully interacting system. 

\begin{figure}[t]
\centering
\includegraphics[width=\columnwidth]{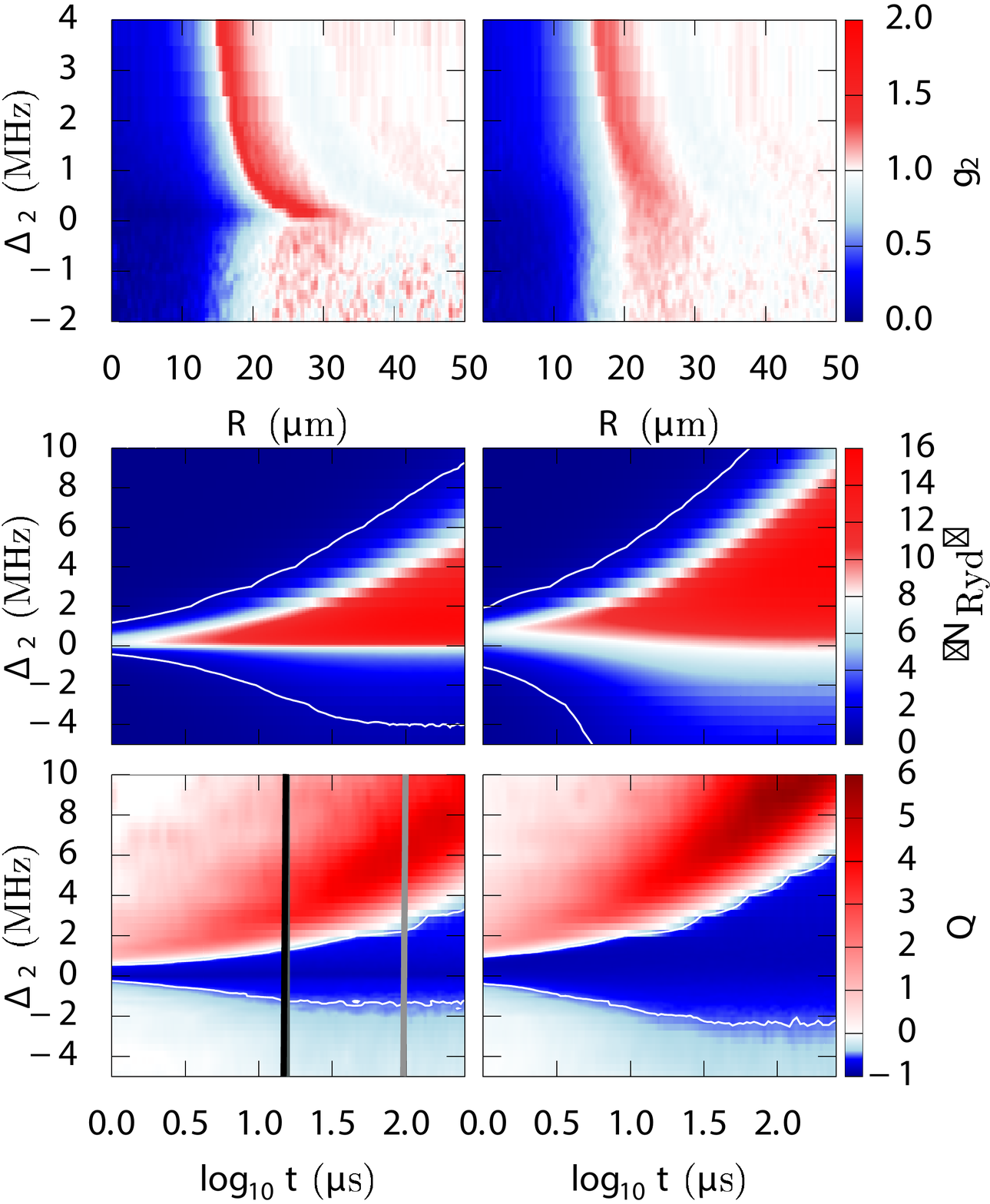}
\caption{\itshape
Statistical measures of Rydberg excitation for a Gaussian profile red 
laser beam (left) and a uniform square laser beam (right).
Top row: $g_2$ map as function of detuning $\Delta_2$ and relative position $R$. In both cases we observe clear blockade for small 
distances $R$ followed by a uniform distribution for negative blue detuning 
($\Delta_2<0$) and strong facilitation for positive blue detuning. 
Middle row: Average Rydberg count across all realizations as function of detuning $\Delta_2$ and time $t$. White contours mark 
$\langle N_\textrm{Ryd}\rangle=1$. Large Rydberg populations do not appear for large 
negative detuning and appear explosively but delayed for positive detunings. 
Note that the square geometry has about half the number of atoms in the 
excitation volume and hence twice the Rydberg fraction.
Bottom row: Mandel $Q$-parameter as function of detuning $\Delta_2$ and time $t$.
Dark blue is $Q<-1/2$,
the deeply subpoissonian regime. For positive detuning $Q>0$, 
we are in the superpoissonian regime (red). Two different times are indicated by vertical lines, 15$\mu$s (black) and 106$\mu$s (gray), for a direct comparison to the same times in Fig.~\ref{fig:expRes}.
\label{fig:GauSqu}}
\end{figure}

On time scales on the order of the steady-state equilibration time for 
a single atom $T=4\Gamma_e/\Omega_1\Omega_2$, 
the system reaches a $1-e^{-1}$ fraction of its final 
Rydberg population, but this is distributed over many single atom 
excitations. At such small time scales, only the edge atoms have a 
larger than average Rydberg probability, as they only have neighbors 
on one side, see Fig.~\ref{fig:mPos}. 
On time scales of $3T$, we observe the first formation of small local 
crystalline structures with correlation lengths larger than 1 lattice 
constant and growing as $t^{1/3}$, see Fig.~\ref{fig:g2map}, consisting 
of two Rydberg excitations 
separated by a single unexcited lattice site. 
The second-order spatial correlation function $g_2$ shown in this figure, and also later in Fig.~\ref{fig:GauSqu}, is a measure for the correlations in the distances between two Rydberg atoms. It can be found by binning the relative distances of Rydberg pairs, and counting the number of these pairs for these binned distances. This distribution is then normalized with respect to the distribution of Rydberg pairs between uncorrelated simulations. For the 1D lattice, these bins are naturally provided by the discrete separation distance measured in lattice sites.
These structures are not all 
consistent with a global crystal, as they have formed at random positions 
in the lattice and could lead to domains in the final state. 
At this time, the enhanced Rydberg probability of the edge sites leads 
to suppression of the Rydberg probability of their neighbors. This 
effect is the beginning of the global crystal structure.

These small crystals will continuously form and melt in the lattice at 
random positions, but as time passes, fewer and fewer sites not consistent 
with a larger crystal will be available, and at time scales of $10T$ 
the average crystal formation will contain three Rydberg atoms. 
As the process continues, the crystal forming on the edge grows, as there 
is no room for excitation hopping, and the average crystal size increases. 
At large time scales ($\sim 100T$), a global crystal has formed by 
spanning the entire lattice, and the correlation length does not increase 
further.

We move on to our quasi 2D random gas, which we will use to model the 
conditions in a magneto optical trap (MOT), and consider three 
statistical properties:  
Average Rydberg count $\langle N_\textrm{Ryd}\rangle$, Mandel 
$Q$-parameter and second-order spatial correlation function $g_2$.

We perform each Monte Carlo realization with a total simulation time 
$t_\textrm{end}=250\,\mu$s, a total number of atoms $N_\text{tot}$ equal 
to the integral of the atomic density in the laser volume. The total 
number of realizations is 2500 for every set of parameters.
Additionally, we assume the gas to be cold enough that we can ignore all 
atomic motion.

\begin{figure*}[t]
\centering
\includegraphics{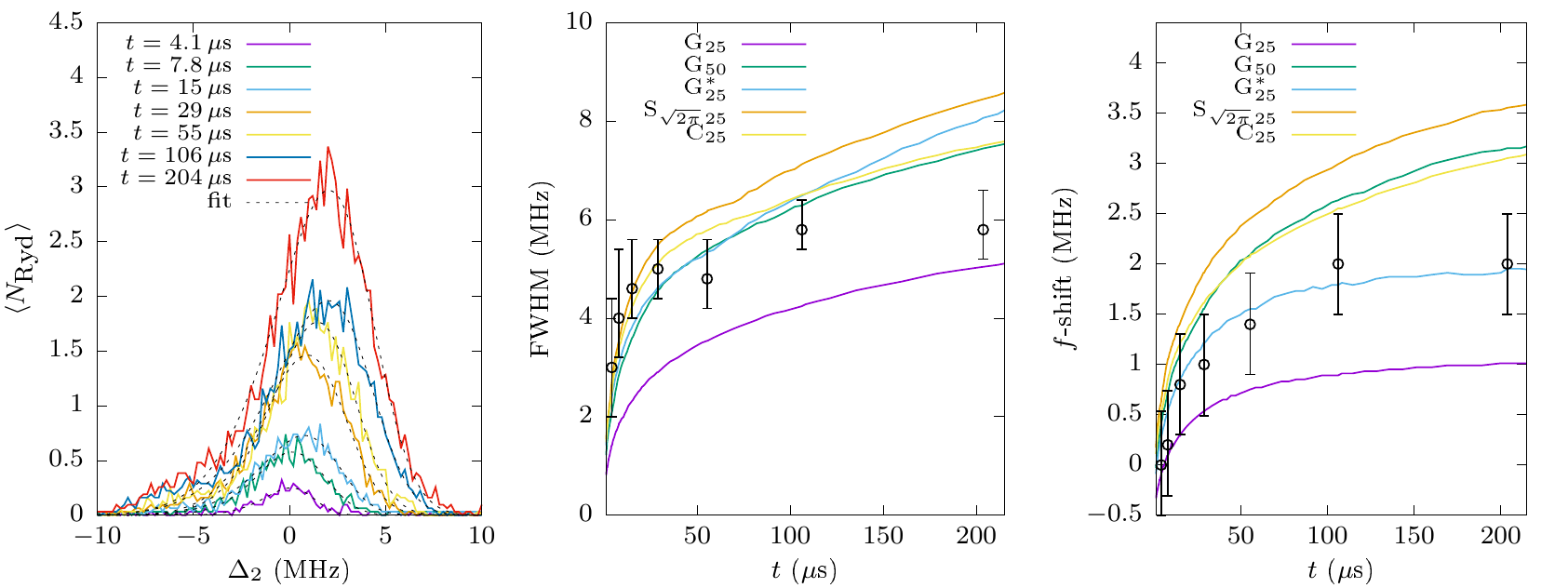}
\caption{\itshape
Left:
Rydberg spectra in experiment (solid) with laser 
parameters $\Omega_1=2\pi\times 4$ MHz, $\Omega_2=2\pi\times 0.5$ 
MHz and $\Delta_1=-2\pi\times 9$ MHz and the red Gaussian profiled laser,
with waist $\sigma_\textrm{red}= 25\,\mu$m, intersects a Gaussian profile 
blue sheet with waist $\sigma_\textrm{blue}= 7\,\mu$m, corresponding to
the simulations in Fig.~\ref{fig:GauSqu}. Dashed curves are skew-Gaussian 
fits.
Middle: Full width half max of the experimental spectrum (circles), 
errorbars determined from fit, and simulation (solid) of 
Gaussian (G), square (S) or circular (C) laser beam profiles with the 
characteristic length scale (std for Gaussian, side length for square 
and radius for circular) denoted by subscript. The star ($*$) denotes 
simulation with twice the red laser intensity 
($\Omega_1\rightarrow \sqrt{2}\Omega_1$).
Right: Frequency shift of the experimental spectrum (circles), 
errorbars determined from fit and frequency drift of the blue laser.
Solid curves represent the same simulations as in middle plot.
The overall comparison between experiment and simulation show 
general agreement for both FWHM and frequency shift. 
\label{fig:expRes}}
\end{figure*}

We excite to the $99S$ state and use laser parameters according to our 
experimental setup: 
$\Omega_1=2\pi\times 4$ MHz, $\Omega_2=2\pi\times 0.5$ 
MHz, $\Delta_1=-2\pi\times 9$ MHz and the blue laser detuning 
$\Delta_2$ is variable. The red laser intersects a Gaussian 
blue sheet with waist $\sigma_\textrm{blue}= 7\,\mu$m. 
Due to the thickness of the blue sheet, we do not expect the 
second order correlation function $g_2$ to be zero inside the blockade 
radius, but strongly suppressed, since we only explore the correlations 
in the projection on the blue laser plane.
In Fig.~\ref{fig:GauSqu}, we show results from two different red 
laser geometries, realizable in our experiment by means of a spatial 
light modulator, a Gaussian beam profile with waist 
$\sigma_\textrm{red}= 25\,\mu$m (left in Fig.~\ref{fig:GauSqu}) and a 
square beam profile of uniform intensity with side length 
$l=\sqrt{2\pi}\,25\,\mu$m (right in Fig.~\ref{fig:GauSqu}). These 
shapes ensure the two lasers output the same power, but the Gaussian 
excitation volume has about twice the number of atoms compared to the 
square geometry. Our model is not limited to these laser shapes and 
parameters, but they show the essential features.
Since the blockade radius $R_\textrm{block}$, for the given parameter 
and $\Delta_2=0$, is $20.9\,\mu$m, we expect the system to be completely 
filled at $N_\textrm{Ryd}=(l/R_\textrm{block})^2\approx 9$ for the 
square beam profile and slightly before that for the Gaussian beam 
profile, we call this number the jamming count $N_\textrm{Jam}$.

We start by generating $N_\text{tot}$ random 3D 
coordinates in the laser volume, and we calculate the laser parameters 
at each coordinate as well as the distances between all pairs of atoms. 
With these parameters in place at the beginning of each realization it is 
easy to evaluate the (de-)excitation rates at each atom on the fly.

For any specific realization we are only interested in the Rydberg count 
and distribution at a number of prespecified time steps, therefore we 
only carry the binary information of Rydberg state or ground state for 
each atom, as well as the atom positions. This will let us determine 
both the Rydberg-Rydberg 
interaction strength at each coordinate from the predetermined atomic 
distances, for use in determining the rates, and is sufficient to 
calculate the aforementioned statistical measures we are interested in 
for the total ensemble of realizations, see Fig.~\ref{fig:GauSqu}.

Analyses of the Monte Carlo simulations show features usually associated 
with the Rydberg-Rydberg interaction, but also illustrates a clear 
dependence on excitation geometry. The location of the first excitation 
is much more likely to occur near the center of the excitation volume 
for the Gaussian profile beam. For negative detunings this leads to 
even stronger blockade as not only is the effective detuning larger, but 
the Rabi frequency is also lower. For positive detunings, however, the 
facilitation rings become narrower. The gradient of the laser intensity 
also leads to a tighter distribution of excitations, as excitation too far 
from the center is unlikely, limiting the number of excitations in the 
volume.

From the Mandel  $Q$-parameter, see Fig.~\ref{fig:GauSqu} (bottom),
we can identify three regimes of interest: 
Firstly, the (weakly sub)poissonian (light blue) regime where $-1/2<Q<0$, 
found for very low blue laser detunings $\Delta_2$. Secondly, the deeply 
subpoissonian (dark blue) regime where $Q<-1/2$, found for small absolute 
values of $\Delta_2$. And thirdly, the superpoissonian (red) regime where 
$Q>0$, found for large positive $\Delta_2$. 

For negative blue laser detunings, the Mandel $Q$-parameter gradually 
decreases over time from 0 to its final value. This happens as atoms 
are excited to the Rydberg state and exclude parts of the volume. 
For very negative detunings, only a few Rydberg excitations exist at any 
given time and the $Q$-parameter stays relatively high, since the jamming 
limit is never reached. This is again reflected in the average Rydberg 
count, which is very low compared to the jamming count $N_\textrm{Jam}$.

For detunings closer to zero, the number of Rydberg excitations 
increases over time and the jamming limit is reached, resulting in 
deeply subpoissonian counting statistics. The subpoissonian regime is 
reached somewhat before the jamming count, since the reduction in the 
excitation volume is significant when 
$N_\textrm{Ryd}\geq\sqrt{N_\textrm{Jam}}$. 

For positive blue laser detunings, an initial Rydberg excitation, called 
the seeding excitation, results in a ring of resonant excitation around 
the seed at the distance $R=(C_6/\Delta_2)^{1/6}$, called the 
facilitation distance $R_\textrm{fac}$. The seeding excitation occurs 
with low probability 
for large detunings, but after seeding more Rydberg atoms are quickly 
excited on resonance. This results in superpoissonian counting 
statistics, as a cascade of Rydberg excitations spreads from the seed. 
We can observe this in the steep slope of the average Rydberg count for positive 
detuning coinciding with very large Mandel $Q$-parameter. 
However, the system quickly fills up, reaching the jamming limit 
resulting in a drop to negative $Q$ and the strongly subpoissonian 
regime.

The second order correlation functions $g_2$ at time $t=250\,\mu$s for 
the Gaussian (left) and square (right) beam profiles are seen in 
Fig.~\ref{fig:GauSqu} (top). 
For negative blue laser detuning, the region in the immediate vicinity 
of a Rydberg excitation ($R\lesssim R_\textrm{block}=21\,\mu$m) shows 
very reduced values of $g_2$. Around $R_\textrm{block}$, the $g_2$ 
gradually climbs to 1, with only a slight overshoot. This behavior is 
the same for both geometries and consistent with the blockade effect. 
The nonzero value for short distances is due to the thickness 
of the excitation volume, as we only consider the correlations in the 
plane parallel to the blue laser sheet.

For positive blue laser detuning the blockade effect is still clearly 
visible for small distances, but at the facilitation distance
$R_\textrm{fac}\approx(C_6/\hbar\Delta_2)^{1/6}$ 
there is a strong signal from the facilitation region followed by a dip 
from the blockade effect of the facilitated excitations. This feature is 
significantly sharper for the Gaussian geometry consistent with a 
narrower facilitation region due to the drop off in laser intensity. 
At about $2R_\textrm{fac}$, a faint signal from the secondary 
facilitation peak is visible for both the Gaussian and the square beam 
profiles.

At very limited blue laser detunings ($|\Delta_2|\lesssim 0.5$ MHz) the 
$g_2$-function resulting from the Gaussian beam profile shows a cusp that 
is not present in case of a square beam profile. This cusp is in part
due to the sharpening of the facilitation peak in the Gaussian profile 
case and in part due to the tighter packing of Rydberg excitations for 
the Gaussian beam profile. This leads to a crystalline locking of the 
Rydberg excitations in the relatively small volume of peak laser 
intensity. 

\section{Experimental comparison}
\label{sec:experiment}

Rydberg excitation was studied experimentally using a setup described 
previously \cite{Engelen14,Bijnen14}. In short, $^{85}$Rb atoms are 
trapped and cooled in a standard magneto-optical trap, resulting in 
typical atomic densities of $10^{16}/$m$^3$ and temperatures of 
$0.2$ mK. To create Rydberg atoms from the cooled sample, the 
780nm trapping laser beams are suddenly switched off, after which a 
separate 780 nm and a 479 nm laser beam are flashed on for a 
variable amount of time, which drive the 
$5S,F=2 \rightarrow 5P_{3/2},F=3$ and $5P_{3/2},F=3 \rightarrow 99S$ 
transitions in $^{85}$Rb. The red laser beam is referenced to 
the atomic transition frequency by a saturated absorption scheme, 
and detuned approximately 9 MHz below resonance. The frequency of 
the blue laser can be scanned in a range of tens of MHz centered 
on the two-photon resonance condition and is referenced to a 
commercial ultrastable cavity (Stable Laser Systems). The 
linewidths of the two laser beams are below 1 MHz but otherwise 
not accurately known.

The red laser beam can be spatially shaped using a spatial 
light modulator \cite{Bijnen14} in various ways but in the 
experiments reported here the spatial shape was a single 
Gaussian with a rms radius of $25\,\mu$m. This shaped red 
beam crosses the blue beam at the center of the MOT, where the 
rms sizes of the blue beam are $7\,\mu$m $\times 1.8$ mm. Typical 
excitation times are in the $\mu$s range. The powers of the laser 
beams were adjusted to provide nominal Rabi frequencies of 
$\Omega_1 \approx 4$ MHz and $\Omega_2 \approx 0.5$ MHz.

Rydberg atoms created by this excitation sequence were detected 
using field ionization. An electric field of several kV/m 
strength is turned on which ionizes any Rydberg atoms present 
and pushes the resulting ions towards a dual microchannel-plate 
detector (GIDS GmbH) \cite{Engelen14}. The current produced by 
the detector is fed through a transimpedance amplifier and then 
sampled by a digital oscilloscope (Agilent DSO 054A). The integral 
of the digitized signal over a period of the experimental cycle is 
taken as proportional to the number of Rydberg atoms produced.

The experiment does not strickly fullfil all conditions assumed in the model.
{\it E.g}., in our model we assume the frozen gas limit to hold. The experiment operates at a temperature of $T=0.2$mK, so that atoms travel the average separation in 50$\mu$s. The time scale set by the rates in Fig.~\ref{fig:rateVar}, however, is only a few microseconds, which then sets the time that an atom spends in a Rydberg excitation. During this time the atoms move over 
a distance of up to a micrometer, which is much smaller than the mean interparticle separation and justifying the frozen gas limit approximation. Note that the coherence time is even smaller, as can be seen from Fig.~\ref{fig:OBEComp}.

Second, for the total decay rate $\Gamma_r$ we took both spontaneous emission and losses from black-body radiation into account. Although this gives rise to a correct lifetime of the Rydberg state, the inclusion of black-body radiation, which dominates the lifetime, is not fully consistent as it does not provide a direct decay to the intermediate state but rather a transition to nearby excited levels. However, these levels will eventually also decay. Also the expression for the van der Waals interaction does not hold up exactly at short distances as other levels are getting nearby. For n=99 the closest pair is 99$P_{3/2}$+98$P_{3/2}$ at an energy of 215 MHz \cite{ARC17}, which is equivalent to the van der Waals energy for two n=99 pairs for a spacing at 8 $\mu$m. However, the repulsive character is maintained, and at this distance the particles are already deeply in the blockaded regime.

Experimentally observed spectra are shown in Fig.~\ref{fig:expRes} 
(left), and fitted with a skew Gaussian. From this fit the derived 
parameters of full width half max (FWHM) (middle) and frequency 
shift ($f$-shift) (right) are determined and plotted (circles). 
The errorbars on the FWHM values 
are determined from the 95\% confidence limit of the fitting 
parameters (fitting error). The $f$-shift errorbars are 
determined as the root of the sum of the fitting error squared and 
the measurement error squared. 
The derived parameters from the experiment are compared to simulation 
(solid). We show here derived parameters for both the Gaussian 
and the square laser profile simulations of Fig.~\ref{fig:GauSqu} 
as well as those from a simulation with the same Gaussian laser 
profile, but twice the intensity, {\itshape ie.}
$\Omega_1\rightarrow \sqrt{2}\Omega_1$. 

Direct comparison between the experimental (see Fig.~\ref{fig:expRes}, 
left) and simulated (Fig.~\ref{fig:GauSqu}, middle) spectra shows 
a general agreement. 
Far from resonance ($|\Delta_2|\gtrsim 8$ MHz) the experimental 
Rydberg count is suppressed at all times. For negative 
Rydberg detuning ($\Delta_2\lesssim-2$ MHz) we see suppression of 
the increase in Rydberg count over time, 
consistent with the existing Rydberg atoms in the volume blocking 
excitation of additional Rydberg atoms. Nearing zero detuning 
($|\Delta_2|\lesssim 2$ MHz) the Rydberg count grows fast and the 
peak shifts towards higher detunings with time.
This means that, especially at later times, the peak is shifted 
to large Rydberg detuning  ($\Delta_2\gtrsim 2$ MHz). 
At these large detunings, we observe almost no Rydberg atoms at 
small times, but with increasing time this changes as facilitation 
shifts the Rydberg levels of unexcitated atoms into resonance.
This is all in agreement with the simulation results presented in 
Fig.~\ref{fig:GauSqu}.

For both experimental data and simulation we observe that both the
FWHM and $f$-shift, derived from the spectra in 
Fig.~\ref{fig:expRes}, values rise quickly and then level out 
after about 50$\,\mu$s. 
The precise shape of the excitation volume and laser profile 
have little influence on the behavior of the FWHM value, and we can 
generally explain experimental observations without knowledge 
of the exact excitation volume geometry. 
Similarly, the $f$-shift shows some dependence on the laser profile, 
but this can be explained from the small dependence of Rydberg count 
on geometry to influence the $f$-shift.

\section{Conclusions}
\label{sec:conc}
We have derived rate equations for excitations of the Rydberg state 
in three-level atoms starting from the master equation. Our approach 
does not assume vanishing populations in the intermediate state and are 
therefore valid for a wide range of laser parameters, in principle 
whenever the three-level approximation of the atom is valid. 

Our rate equation model agrees with the master equation, 
provided that only one eigenvalue has not dampened out. We have 
investigated and described the validity range of our approach, 
and determined criteria that provide 
sufficient insight into whether our model is valid or full solution of 
the master equation is required.

We have made a Monte Carlo implementation of our (de-)excitation rates, 
and explored different excitation geometries and laser parameters. In 
this paper we have reported on 1D lattice simulations, with 
parameters corresponding to previous theoretical work published in 
\cite{PhysRevA.87.023401}, and found our results to be consistent with 
literature. We explored the dynamics of self assembly of the resulting 
1D Rydberg crystal, and the time evolution of the second order 
correlation function $g_2$ and site dependent excitation probability. 

We further explored 2D settings, where we considered the effect of the 
laser beam profile by comparing a Gaussian profile to a square of 
uniform intensity with sharp edges. We found that the beam profile has 
significant influence on the resulting excitation pattern and that a 
Gaussian profile in general will result in sharper features in 
the $g_2$-map, but at the cost of lower excitation counts. 

We have compared our model to experimentally observed Rydberg 
spectra at several time steps and found a general agreement for 
the spectral shapes and derived parameters FWHM and $f$-shift. We 
did not see any significant dependence on excitation volume geometry 
in the time dependence of FWHM, but the geometry 
dependence of the Rydberg count may result in a slight geometry 
dependence of the $f$-shift.

\begin{acknowledgments}
This research was financially supported by the Foundation for 
Fundamental Research on Matter (FOM), and by the Netherlands 
Organization for Scientific Research (NWO). We also acknowledge 
the European Union H2020 FET Proactive project RySQ (grant N. 
640378).
\end{acknowledgments}

\bibliography{threeLevelRateEquations_arXiv}

\appendix

\end{document}